# Stable, high-performance operation of a fiber-coupled superconducting nanowire avalanche photon detector

SHIGEHITO MIKI,[1,2,*] MASAHIRO YABUNO,[1] TARO YAMASHITA,[1] AND HIROTAKA TERAI[1]

[1]*Advanced ICT Research Institute, National Institute of Information and Communications Technology, 588-2 Iwaoka, Nishi-ku, Kobe, Hyogo 651-2492, Japan*
[2]*Graduate School of Engineering Faculty of Engineering, Kobe University, 1-1 Rokkodai-cho, Nada-ku, Kobe-city, Hyogo 657-0013, Japan*
*\*s-miki@nict.go.jp*

**Abstract:** Recent progress in the development of superconducting nanowire single photon detectors (SSPD or SNSPD) has delivered excellent performance, and has had a great impact on a range of research fields. Significant efforts are being made to further improve the technology, and a primary concern remains to resolve the trade-offs between detection efficiency (DE), timing jitter, and response speed. We present a stable and high-performance fiber-coupled niobium titanium nitride superconducting nanowire avalanche photon detector (SNAP) that resolves these trade-offs. We demonstrate afterpulse-free operation in serially connected two SNAPs (SC-2SNAP), even in the absence of a choke inductor, achieving a ~7.7 times faster response speed than standard SSPDs. The SC-2SNAP device showed a system detection efficiency (SDE) of 81.0% with wide bias current margin, a dark count rate of 6.8 counts/s, and full width at half maximum timing jitter of 68 ps, operating in a practical Gifford–McMahon cryocooler system.

## 1. Introduction

Recent progress in the development of the superconducting nanowire single photon detector (SSPD or SNSPD) has delivered system detection efficiency (SDE) approaching unity, a low dark count rate (DCR), and short timing jitter [1-3]. This has had a great impact on a number of research field, including quantum information, laser communications, laser sensing, and fluorescent observation [4-7]. Significant efforts are being made to achieve further improvement, and a primary concern at present is to resolve the trade-off between different specifications. In particular, the trade-off between detection efficiency (DE), timing jitter, and response speed is yet to be resolved. For example, NbTiN based SSPDs have shown SDEs as high as ~80% at telecommunication wavelengths [2], but do not perfectly saturate in bias current dependency, implying that internal DE is not reached to 100%. An effective way to achieve 100% internal DE is to create a narrower nanowire [8]. This, however, makes the switching current ($I_{sw}$) of the nanowire smaller, degrading the timing jitter by reducing the signal-to-noise ratio (SNR) of the output. It also requires a longer nanowire to cover a sensitive area of the same size, resulting in larger kinetic inductance and hence slowing the response speed. A higher internal DE could also be achieved by using a superconducting material with a lower superconducting gap, but this also results in reducing the $I_{sw}$, and degrades the timing jitter [3].

The trade-off between the sensitive area of the device and the response speed has also yet to be resolved. In applications such as free- space laser communication and fluorescent correlation spectroscopy (FCS) [5, 7], a large sensitive area is needed to couple efficiently with incident light from free space, or a multimode (MM) fiber with a larger core size than a single mode fiber. However, enlargement of the sensitive area is accompanied by a large kinetic inductance, resulting in a slow response speed.

The superconducting nanowire avalanche photon detector (SNAP) is an alternative configuration that is able to resolve these trade-offs [9]. SNAP basically comprises N nanowire sections placed in parallel, with a serially connected choke inductor to confine the bias current to the device side, allowing avalanche switching. Three-dimensional 2SNAPs based on WSi nanowires, in which two nanowire sections are placed in parallel, have shown detection efficiencies as high as ~87%, with a ~2 higher SNR than standard SSPD [10]. However, a large inductance choke inductor is necessary for avalanche operation, precluding high-speed operation. Serially connected 2SNAP (SC-2SNAP) is an alternative structure that can operate without a choke inductor [9], and for which saturated internal DE over a wide bias current margin has been reported [11]. However, its sensitive area is relatively small (~2 × 20 μm$^2$) and a choke inductor is still used. In this work, we report the development of an NbTiN-2SNAP with a 15 × 15 μm$^2$ sensitive area size for efficient optical coupling with a single mode (SM) optical fiber. We demonstrate the effectiveness of SC-2SNAP by comparing it with a standard 2SNAP (ST-2SNAP). We also examine its stable operation without afterpulsing, as the stable operation of SNAP has been raised as a critical issue in many applications [12]. As an effective and convenient way of testing stable operation, we employed an autocorrelation measurement. This can also be used to evaluate the dead time of the detector. We then demonstrated the overall performance in terms of SDE, system dark count rate (SDCR), and timing jitter.

## 2. Experimental procedure

For fabrication of the 2SNAP devices, 7.5-nm-thick NbTiN thin films were prepared on Si substrates with a 240-nm-thick thermally oxidized $SiO_2$ layer by DC magnetron sputtering [13]. The NbTiN films were then formed to nanowire structures by e-beam lithography and reactive ion etching (RIE). We fabricated two different structures, as shown in Fig. 1. ST-2SNAP comprised two nanowire meandering sections ($7.5 \times 15$ μm$^2$ area size each) placed in parallel, across a sensitive area of $15 \times 15$ μm$^2$. In SC-2SNAP, several tens (depending on the nanowire width and space between nanowires) of parallel pairs of 15-μm-long straight nanowires were connected in series, covering the sensitive area. A choke inductor comprising a 300-nm-wide superconducting meandering line covering the $15 \times 60$ μm$^2$ area was formed in the same NbTiN layer in several of the devices, between the sensitive area and a 100-nm-thick NbN coplanar waveguide line. The $I_{sw}$ of the choke inductor was sufficiently large as a wider line width was used than in the sensitive area so not to react against single photon incidents. An optical cavity structure comprising a 250-nm-thick SiO dielectric layer and 100-nm-thick Ag mirror was placed on to the sensitive area to enhance absorptance of incident photons at a wavelength of 1550 nm [1]. We fabricated a number of 2SNAP designs, and conventional SSPD devices for comparison. Table 1 lists the designs (types, sensitive area size, nanowire width, and space between nanowires) and physical parameters of these devices (inductance of the device ($L_d$), inductance of the choke inductor ($L_s$), critical temperature ($T_c$), and switching current of the device ($I_{sw}$)). The $L_d$ was obtained by measuring the phase of a reflected microwave signal against the frequency, using a network analyzer [14]. The $L_s$ was then estimated from the $L_d$ and device design. Fabricated devices were mounted into compact packages, allowing efficient optical coupling between the device sensitive area and incident light from a single mode optical fiber with GRIN lenses [15]. The packaged devices were then installed into a Gifford–McMahon cryocooler system with an operational temperature of ~2.2 K. The procedures for measuring the photon response count rate, to ensure detection efficiency in the 1550 nm wavelength region, dark count rate, and timing jitter were as described in previous papers [1].

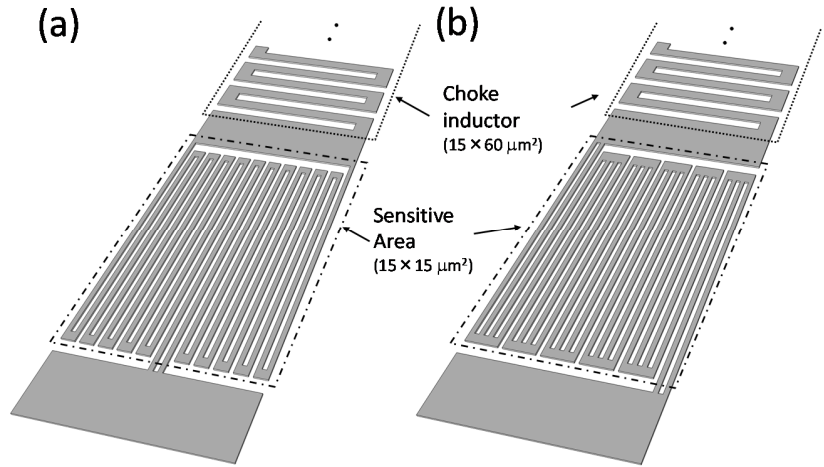

Fig. 1. Schematic view of structure of (a) ST-2SNAP and (b) SC-2SNAP.

**Table 1. Design and physical parameters for devices treated in this paper**

| # | Type | Width/space (nm) | $L_d$ ($\mu H$) | $L_s$ ($\mu H$) | $T_c$ (K) | $I_{sw}$ ($\mu A$) |
|---|------|------|------|------|------|------|
| 1 | ST-2SNAP | 80/80 | 1.06 | 0.50 | 7.8 | 25.1 |
| 2 | SC-2SNAP | 80/80 | 1.08 | 0.51 | 7.8 | 24.4 |
| 3 | SC-2SNAP | 60/80 | 0.76 | none | 7.4 | 19.5 |
| 4 | SSPD | 60/80 | 2.66 | none | 7.5 | 10.6 |

## 3. Experimental Results

### 3.1 Comparison between ST- and SC-2SNAP

We first investigated the photon response count rate of the ST- and SC-2SNAP and compared their bias current dependencies, to gain an understanding of the avalanche operation mechanism in devices with a large sensitive area. To make the comparison fair, we chose device #1 and #2 from Table. 1, as these had identical nanowire width, space, and $L_s$. Figure 2 (a) shows the bias current dependencies normalized by $I_{sw}$ on a photon response count normalized by the counts at the saturated region, against the incident photon flux of the $10^6$ photon/s at the input port of the cryocooler system. Although SNAP devices should ideally work under avalanche operation, in which cascade switching in all the nanowire sections happens after triggering of the initiating section, they can also work in arm-triggered operation where only the initiating layer is triggered [12]. Since the output pulse height produced by an arm-trigger operation is much smaller than that produced by an avalanche operation, the threshold level of the counter is set sufficiently high as to exclude the signals from the arm-triggered operation. Both devices responded to incident photons and showed saturated count rates at high bias current regions of above ~0.8 $I_{sw}$. The bias current dependencies in the low bias current region were different for the different structures; ST-2SNAP showed an abrupt increase in photon response at a bias current of 0.75–0.85 $I_{sw}$, whereas SC-2SNAP showed a photon response even at low bias currents, increasing monotonically as the bias current increased.

To clarify the difference in bias current dependency of the count rate between ST- and SC-2SNAP, we simulated the bias current flow of 2SNAP using the equivalent circuit shown in Fig. 2 (b), by applying a SPICE model. In this circuit, 2SNAP devices were represented by the choke inductor and the sensitive area comprising the initiating section A, avalanche section B, and non-reactive sections C. The temporal transition to the resistive state in initiating section A could be expressed using the well-known equivalent circuit comprising $L_{secA}$, $R_N$, and a switch [16]. Only $L_{secB}$ and $L_r$ were placed in sections B and C, respectively, to simplify the circuit model, and $L_r$ was set to zero in the case of ST-2SNAP. We used this model to examine the current flowing to avalanche section B ($I_{secB}$) after the transition to the resistive state in initiating section A. Avalanche switching in section B occurs if $I_{secB}$ exceeds $I_{sw}$', where $I_{sw}$' is the switching current of section A and B and is 0.5 $I_{sw}$, on the assumption that the two sections have the same switching current. The minimum bias current needed to trigger avalanche switching was defined as avalanche current ($I_{AV}$), and verified for each structure. From the simulations, $I_{AV}$ for ST-2SNAP with design#1 was estimated to be 0.74 $I_{sw}$, which can explain the bias current dependencies shown in Fig. 2(a). Though it is possible to realize a lower $I_{AV}$ in ST-2SNAP by using a choke inductor with a larger $L_s$, this results in a slower response speed. Meanwhile, the $I_{AV}$ for SC-2SNAP with design #2 was estimated to be 0.52 $I_{sw}$, which is significantly lower than that for ST-2SNAP, producing an avalanche operation even in the low bias current region. This could be explained qualitatively if the $L_{secA}$ is much lower than that for ST-2SNAP, and $L_r$ works effectively as a choke inductor, avoiding a leakage of current to the load side. It should be noted that the estimated $I_{AV}$ did not change significantly, even in the absence of $L_s$. This simulated result supports a suggestion in [9] that an additional choke inductor to limit the response speed is unnecessary for an SC-2SNAP with a relatively large sensitive area.

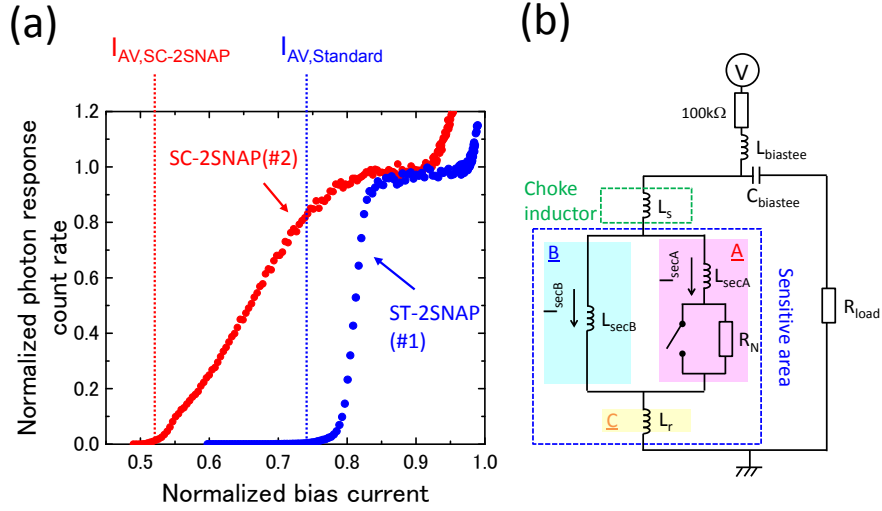

Fig. 2. (a) Normalized photon response count rate as a function of normalized bias current for SC-2SNAP and ST-2SNAP. (b) Equivalent circuit of 2SNAP used for SPICE simulation.

*3.1 Stable and high speed operation of SC-2SNAP*

Since the favorable features of SC-2SNAP were confirmed by the photon response measurements, we next investigated the stability of operation by measuring the autocorrelation function $G(\tau)$. $G(\tau)$ is the average of a temporal correlation between two signals with interval $\tau$, and is widely used for the observation of the correlated characteristics, for example in FCS experiments [7]. If there is no deviated autocorrelation in incident photons, such as from a continuous wave (CW) laser source, $G(\tau)$ reflects the characteristics of the detector itself. Figure 3 (a) shows the photon response count rate as a function of the bias current for SC-2SNAP without $L_s$ (device #3), using a CW laser source, and (b), (c), and (d) show three $G(\tau)$ curves at different bias currents of 12.0, 17.0, and 18.4 µA, respectively. For comparison, $G(\tau)$ curves for a standard SSPD with the same nanowire design (#4), and an SC-2SNAP with a choke inductor (#2), are shown in Fig. 3 (c). All the $G(\tau)$ curves for #3 at the three bias currents showed sharp drops to zero in the temporal range of around 10–100 ns, which reflect the dead times of the detector. These dead times appeared more quickly as the bias current was increased, as shown in the figures. This is due to the faster recovery of DEs as the bias current dependency approaches the saturated region [16]. The dead time at 17.0 µA was 20 ns, where the dead time was defined as the time at which $G(\tau)$ drops to 0.9. This was clearly faster than the standard SSPD #5 (153 ns) and SC-2SNAP with choke inductor #2 (70 ns), because of the smaller $L_d$.

To identify correct and stable operation without afterpulses, the flat $G(\tau)$ curve must be within the temporal range above the dead time of the detector. Device #3 showed a slight deviation in the curves in the temporal range 100 ns–10 µs, at a bias current of 12.0 and 18.4 µA. A flat $G(\tau)$ curve was obtained at a bias current of 17.0 µA, suggesting no characteristic auto-correlation in the operation of the detector, and no afterpulses. We also confirmed the flat $G(\tau)$ curves in a bias current region of 15–18 µA, which is the preferred region with a high internal DE and shown in pink in Fig. 3 (a).

We next consider the causes of the deviation from a flat curve in $G(\tau)$, at bias currents of 12.0 and 18.4 µA. In SNAP devices, a loss of flatness in $G(\tau)$ may be due to several causes: (i) afterpulsing due to an undesired avalanche operation after the arm-triggered operation in the bias current region lower than $I_{AV}$ [12], (ii) afterpulsing caused by thermal relaxation dynamics [12, 17], and (iii) DE or (iv) DCR fluctuation due to excess current flow to the device during the recovery process, caused by the AC-coupled readout circuit [18]. First, afterpulses due to (i) can be excluded as a possible cause because the bias currents of 12.0 µA and 17.0 µA were both above $I_{AV}$. Indeed, output pulses originating from the arm-triggered operation, which should have a much smaller pulse height than that in the avalanche operation, were not observed. Afterpulses due to (ii) are difficult to classify as a possible cause, because their dynamics usually occur at a much faster time scale of ~a few hundred picoseconds [12]. We attribute the deviation of G(τ) at the low bias current of 12.0 µA to possible cause (iii), because a strong slope in DE dependency can be observed, and a dark count rarely occurs, even if the bias current exceeds its initial value. Meanwhile, at the highest bias current of 18.4 µA, afterpulses due to (iv) are a plausible cause. In this case, an excess current due to the read out circuit could account for the unexpected increase of DCR, resulting in afterpulses. We optimistically suppose that improving the readout circuit, for example by a DC coupled readout scheme, will resolve this effect, by keeping the overshoot effect as low as possible.

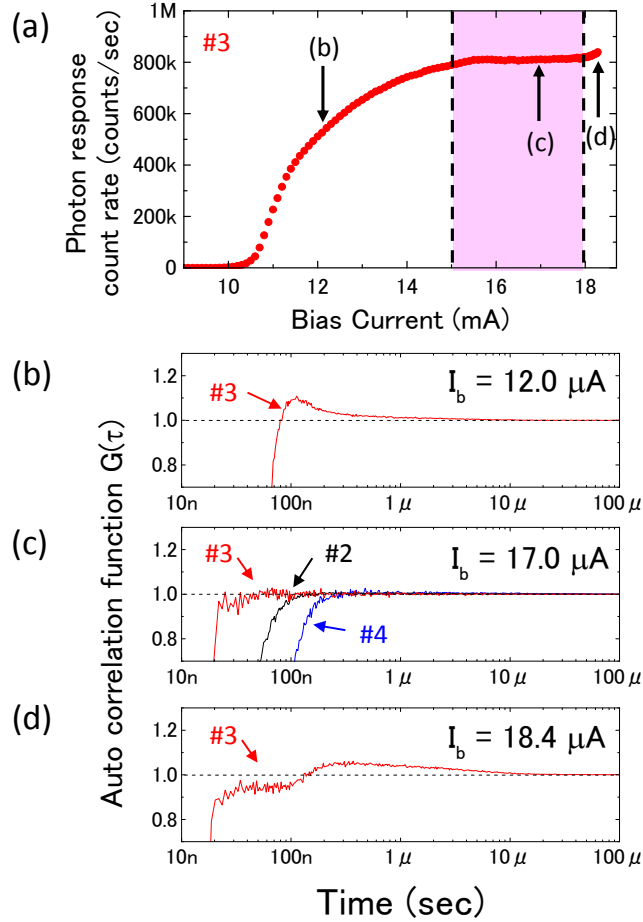

Fig. 3. (a) Photon response count rate as a function of bias current for SC-2SNAP without $L_s$ (#3). (b)(c)(d) Auto correlation function G(t) of #3 at bias currents of 12.0, 17.0, and 18.4 µA, respectively. For comparison, G(τ) of SC-2SNAP with choke inductor (#2) and standard SSPD (#4) are shown as the black and blue curves in (c), respectively.

*3.1 System efficiency, dark count rate, and timing jitter of SC-2SNAP*

We finally discuss the overall performance of the system in terms of the SDE, SDCR, and timing jitter of SC-2SNAP (#3). Figure 4 shows the SDE and SDCR as a function of the bias current in the stable operation range of 15-18 μA, as confirmed in previous section. Therefore, we were able to calculate the SDE without considering the influence of afterpulses by applying the following formula: SDE = ($R_{light}$ − $R_{dark}$)/$R_{incident}$ [2], where $R_{light}$ is the output count rate when the incident photon is irradiated, $R_{dark}$ is the SDCR, and $R_{incident}$ is the incident photon flux rate at the input port of the cryocooler system. As can be seen, the device showed a flat DE with an average value of 81.0% in this range. The flat SDCR dependency in the bias current range of 15-17.5 μA was also shown and average value was 6.8 counts/s, which is considerably lower than the reported values for conventional SSPDs [1, 2]. The full width of the device at half maximum (FWHM) timing jitter was also measured using a laser pulse source with a 100 fs pulse width and standard time-correlated single photon counting (TCSPC) module (Hydra Harp 400, Pico Quant Inc.) [1, 2]. Fig. 5 shows observed FWHM timing jitter as a function of bias current in the range of stable operation. The timing jitter decrease with increasing bias current due to the increase of SNR, and obtained 68 ps at the bias current of 17.5 μA at which the obtained histogram of timing correlation between laser pulses and output signals is shown in the inset of Fig. 5.

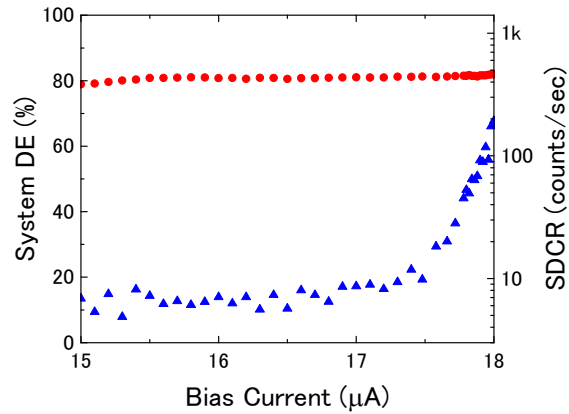

Fig. 4. SDE and SDCR as a function of bias current for SC-2SNAP (#3) within the range of stable operation.

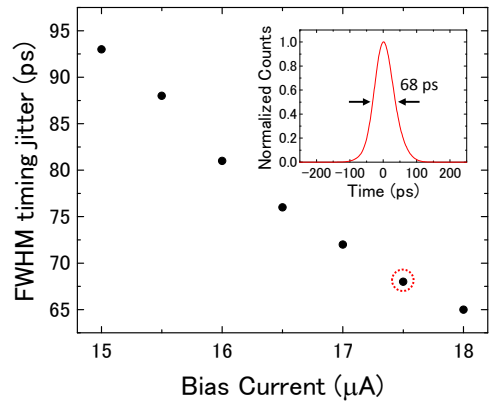

Fig. 5. FWHM timing jitter of SC-2SNAP (#3) as a function of bias current in the range of stable operation. Inset shows the histogram of timing correlation between laser pulses and output signals at the bias current of 17.5 μA, indicating FWHM timing jitter of 68 ps.

## 4. Summary

In summary, we examined the stable operation and system performance of a fiber coupled NbTiN-2SNAP with a 15 × 15 μm$^2$ sensitive area, under GM cryocooler operation. A comparison between the SC-2SNAP and ST-2SNAP confirmed that single photon responses above their own $I_{AV}$ were detected in both devices, though the $I_{AV}$ in SC-2SNAP was lower than that in ST-2SNAP. Autocorrelation measurement revealed that afterpulse-free operation in SC-2SNAP could be realized at a bias range with saturated DE. The stability of this operation was confirmed without requiring the addition of a choke inductor, resulting in ~4 times lower inductance and 7.7 times faster response than conventional SSPDs. The fiber-coupled NbTiN SC-2SNAP showed an SDE of 81.0%, with a relatively wide bias range of 15.0–18.0 μA, a low SDCR of 6.8 counts/s, and timing jitter of 68 ps. The simultaneous achievement of high SDE, low SDCR, and short timing jitter in a GM cryocooler system offers not only high performance but also reliability and convenient handling for users. This will accelerate the adoption of SNAPs in a range of novel applications. The achievement of a wide bias margin in the GM cryocooler system is also an important milestone in the realization of large scale SSPD arrays with near unity efficiency in all pixels.

**Funding**

Japan Science Technology, Core Research for Evolutional Science and Technology (JST-CREST).

**Funding and Acknowledgments**

We thank Dr. Saburo Imamura for his operation in e-beam lithography. We also thank Dr. Johtaro Yamamoto, Prof. Masataka Kinjo, Prof. Yasushi Hiraoka, and Dr. Tokuko Haraguchi for their helpful discussion about auto correlation measurement.